\documentclass[sigconf]{acmart}
\usepackage{multirow}
\usepackage[normalem]{ulem}
\usepackage{algorithm}
\usepackage{algpseudocode}
\useunder{\uline}{\ul}{}

\AtBeginDocument{%
  }

\setcopyright{acmcopyright}
\copyrightyear{2023}
\acmYear{2023}
\acmDOI{XXXXXXX.XXXXXXX}

\acmConference[Conference acronym 'XX]{Make sure to enter the correct
  conference title from your rights confirmation emai}{June 03--05,
  2018}{Woodstock, NY}
\acmPrice{15.00}
\acmISBN{978-1-4503-XXXX-X/18/06}

\begin{document}

\title{Federated Unlearning for On-Device Recommendation}

\author{Wei Yuan}
\affiliation{%
  \institution{The University of Queensland}
  \city{Brisbane}
  \country{Australia}
}
\email{w.yuan@uq.edu.au}

\author{Hongzhi Yin}\authornote{Corresponding author.}
\affiliation{%
  \institution{The University of Queensland}
  \city{Brisbane}
  \country{Australia}
}
\email{h.yin1@uq.edu.au} 

\author{Fangzhao  Wu}
\affiliation{%
  \institution{Microsoft Research Asia}
  \city{Beijing}
  \country{China}
}
\email{wufangzhao@gmail.com}

\author{Shijie Zhang}
\affiliation{%
  \institution{Tencent}
  \city{Shenzhen}
  \country{China}
}
\email{julysjzhang@tencent.com}

\author{Tieke He}
\affiliation{%
  \institution{Nanjing University}
  \city{Nanjing}
  \country{China}
}
\email{hetieke@gmail.com}

\author{Hao Wang}
\affiliation{%
  \institution{Alibaba Cloud, Alibaba Group}
  \city{Hangzhou}
  \country{China}}
\email{cashenry@126.com}

\begin{abstract}
  The increasing data privacy concerns in recommendation systems have made federated recommendations attract more and more attention.
  Existing federated recommendation systems mainly focus on how to effectively and securely learn personal interests and preferences from their on-device interaction data. Still, none of them considers how to efficiently erase a user's contribution to the federated training process.
  We argue that such a dual setting is necessary.
  First, from the privacy protection perspective, ``the right to be forgotten (RTBF)'' requires that users have the right to withdraw their data contributions.
  Without the reversible ability, federated recommendation systems risk breaking data protection regulations.
  On the other hand, enabling a federated recommender to forget specific users can improve its robustness and resistance to malicious clients' attacks.
  
  To support user unlearning in federated recommendation systems, we propose an efficient unlearning method FRU (\emph{F}ederated \emph{R}ecommendation \emph{U}nlearning), inspired by the log-based rollback mechanism of transactions in database management systems.
  It removes a user's contribution by rolling back and calibrating the historical parameter updates and then uses these updates to speed up federated recommender reconstruction. However, storing all historical parameter updates on resource-constrained personal devices is challenging and even infeasible. In light of this challenge,  we propose a small-sized negative sampling method to reduce the number of item embedding updates and an importance-based update selection mechanism to store only important model updates.
  To evaluate the effectiveness of FRU, we propose an attack method to disturb federated recommenders via a group of compromised users. Then, we use FRU to recover recommenders by eliminating these users' influence.
  Finally, we conduct extensive experiments on two real-world recommendation datasets (i.e. MovieLens-100k and Steam-200k) with two widely used federated recommenders to show the efficiency and effectiveness of our proposed approaches.
\end{abstract}



\maketitle

\section{Introduction}\label{sec:intro}
Recommender Systems (RS) suggest the most appropriate items and services to users by analyzing the collected personal data, e.g. user-item interactions and user profiles~\cite{chen2018tada,zhang2020gcn,long2022decentralized}.
With the growing awareness of privacy and the recent publishing of data privacy protection regulations such as the General Data Protection Regulation (GDPR)~\cite{voigt2017eu} in the European Union and the California Consumer Privacy Act (CCPA)~\cite{harding2019understanding} in the United States,
collecting, storing, and using users' data is becoming harder.
To address the above challenges, more and more researchers focus on applying federated learning~\cite{mcmahan2017communication} to recommendation systems (FedRecs), which train recommender models on client devices without sharing user data with a central server or other clients.

Since Ammad et al.~\cite{ammad2019federated} proposed the first federated recommendation framework, FedRecs have made great advancements recently~\cite{imran2022refrs}.
For example, Muhammad et al.~\cite{muhammad2020fedfast} proposed FedFast to accelerate training convergence.
Lin et al.~\cite{lin2020fedrec} investigated how to exploit explicit feedback.
Liang et al.~\cite{liang2021fedrec++} attempted to improve the security of FedRecs via denoising techniques.
Wu et al.~\cite{wu2021fedgnn} incorporated GNN into a general FedRec framework.

Despite great advancements made in this area, it has been unexplored how to forget specific users during the FedRec training process. 
Without the ability to erase specific users' contributions to the federated training process, FedRec might break privacy protection laws or regulations such as CCPA and GDPR that give users the right to control and withdraw their data at any time.
Apart from reducing the risks of breaking privacy protection rules, implementing unlearning is also important to improve FedRec's robustness and resistance to malicious attacks in an open setting where any user/device can participate in the training process of FedRec.
Recent studies~\cite{zhang2022pipattack,rong2022fedrecattack} have shown that current FedRecs are still not ``safe'' enough when facing malicious users' attacks.
After detecting such attacks, the ability to efficiently erase such malicious users' influence without retraining from scratch is essential for FedRecs. 

Although some recent works~\cite{chen2022recommendation,li2022making} tried to apply \emph{machine unlearning} to recommender systems because of data privacy concerns, all of them focus on the traditional centralized recommenders. 
Their methods require accessing the whole training data during unlearning, which is prohibitive in FedRecs.
Some works~\cite{liu2020learn,liu2022right,wu2022federated,liu2021federaser} explored unlearning in federated learning, however, they are tailored for classification tasks in Computer Vision (CV) area.
To erase the contributions of target clients, the most naive yet effective method is to retrain the recommender model from scratch after removing the target clients, which is infeasible in the real-world recommendation setting due to its huge time and resource costs.
Another alternative is to continue the training after removing the target clients.
However, such a method cannot guarantee whether and when these target users' influence on the global parameters (e.g. item embeddings) will be erased.
As a result, how to effectively and efficiently erase target clients' contributions is not trivial in FedRecs.

Inspired by the log-based rollback mechanism of transactions in database management systems (DBMS), we propose to record each client's historical model updates. Once we need to erase some users' contributions, we will roll back and reconstruct the other clients' models according to their training logs (i.e. their historical model updates).
To achieve that, the most intuitive way  is to keep all clients' historical model updates at the central server. 
This method can work when the number of clients is small, such as in the federated classification setting in which there are only tens of clients~\cite{liu2021federaser}.
However, in FedRec, the number of clients is several orders of magnitude larger than the classification settings.
As the storage costs at the central server increase linearly with the number of clients, this naive method is unsustainable and impractical for FedRec.
Therefore, we propose to retain historical model updates at each client's local device, and storage costs at each client device decouple with the number of clients. Still, it is non-trivial to store each client's historical model updates on the resource-constrained device, and it is infeasible to simply store all updates. 

In this paper, we propose FRU (\emph{F}ederated \emph{R}ecommendation \emph{U}nlearning), a simple yet effective federated recommendation unlearning method.
FRU is model-agnostic and can be applied to most federated recommendation systems.
The basic idea of FRU is to erase a target client's influence by revising FedRec's historical updates and leveraging the revised updates to speed up FedRec reconstruction.
Compared with completely retraining (reconstructing) the FedRec  from scratch, FRU requires less running time and achieves even better model performance.
FRU stores each client's historical updates locally on decentralized personal devices to avoid high central server storage overhead. To efficiently utilize the limited storage space on each client's device, we design two novel components: a user-item mixed semi-hard negative sampling component and an importance-based update selection component.
The user-item mixed negative sampling exploits high-quality negative samples to train FedRec, reaching comparable model performance with fewer negative samples than the traditional sampling method.
Consequently, it reduces the size of item embedding updates at each client. 
The importance-based update selection component dynamically chooses important updates to store on each client device at each training epoch, instead of storing all parameter updates.

After achieving unlearning in FedRec, evaluating the effectiveness of unlearning is neither easy because there are a large number of users in recommendation datasets, and  many of them have common items and similar preferences (actually, it is the base of CF-based recommendation methods).  Deleting a small portion of normal users/clients will not significantly change the performance of FedRec. In light of this, we propose an attack method to destroy the FedRec with a group of compromised clients/users (also called malicious users). An effective unlearning method should recover the destroyed FedRec quickly and achieve comparable or even better performance than training without malicious users.

To demonstrate the effectiveness of our proposed approach, we choose two commonly used recommenders~\cite{zhang2021graph}, Neural Collaborative Filtering (NCF)~\cite{he2017neuralcoll} and LightGCN~\cite{he2020lightgcn}, with the most basic federated learning protocol~\cite{ammad2019federated} as our base models.
Then, we conduct extensive experiments with these base models on two real-world recommendation datasets, MovieLens-100k and Steam-200k.
The experimental results show that FRU can erase the influence of removed malicious users, with at least $7$x speedup compared with the naive retraining from scratch.

The main contributions of this paper are summarized as follows:
\begin{itemize}
  \item To the best of our knowledge, this is the first work to investigate machine unlearning in federated recommender systems,  enabling FedRecs to effectively erase the influence of specific users/clients and efficiently finish recovering.
  \item We propose FRU, an unlearning method tailored for FedRecs. It stores each client's historical changes locally on their devices. To improve the storage space efficiency on resource-constrained devices, we propose a novel negative sampling method and an importance-based update selection mechanism. Then, FRU rolls back FedRecs to erase the target users'/clients' influence and fast recover FedRecs by calibrating the historical model updates.
  \item We design an attack method to intuitively evaluate the unlearning effectiveness. The experimental results demonstrate the effectiveness and efficiency of FRU on two real-world datasets. Comprehensive ablation studies reveal the effectiveness and importance of each technical component in FRU.
\end{itemize}

\section{Preliminaries}

\subsection{Federated Recommendation}
Let $\mathcal{V}$ and $\mathcal{U}$ denote the sets of items and users (clients), respectively. The sizes of items and users are $\left| \mathcal{V} \right|$ and $\left| \mathcal{U} \right|$.
Each user $u_{i}$ owns a local training dataset $\mathcal{D}_{i}$, which contains user-item interactions $(u_{i}, v_{j}, r_{ij})$.
$r_{ij}=1$ represents $u_{i}$ has interacted with item $v_{j}$, and $r_{ij}=0$ means no interactions exist between $u_{i}$ and $v_{j}$ (i.e. negative samples).
We denote the set of all negative instances for $u_{i}$ as $\mathcal{V}_{neg}(i)$.
The federated recommender is trained to predict the score of $\hat{r}_{ij}$ between $u_{i}$ and all non-interacted items.
And then, according to the predicted scores $\hat{r}_{ij} \in [0, 1]$, the federated recommender outputs an item list $\hat{\mathcal{V}}_{i}$ for each user $u_{i}$ by selecting top $K$ ranked items w.r.t. $\hat{r}_{ij}$.

With FedRec, users can keep their private data locally and do not need to share it with others.
To achieve this goal, a central server is often employed to coordinate individual clients.
At first, the central server randomly selects a batch of users/clients and dispenses the global parameters to these clients.
Each client combines its received global parameters and its local user embedding as a local recommender. 
Then, the local recommender is optimized on local data using a certain objective function (e.g. BPR loss~\cite{rendle2012bpr}).
After multiple local training epochs, each client sends the updated global parameters (or global parameter updates) back to the server and keeps the updated private parameters locally.
The server aggregates received global parameter updates with a certain aggregation strategy~\cite{mcmahan2017communication}.
The above steps, namely the global training round, are repeated until the system satisfies the pre-defined requirements.

\subsection{Machine Unlearning of Federated Recommendation}\label{sec_pre_mufr}
Machine unlearning (also referred to as data removal or data deletion~\cite{ginart2019making,guo2019certified}, selective forgetting~\cite{golatkar2020eternal}) aims to actively clean the influence of a specific subset of training data.
In this paper, we apply machine unlearning to remove the contributions of a subset of users $\mathcal{U}^{'}$ in FedRecs. 
Specifically, let $f_{ori}$ be the original federated recommender, and after applying unlearning to $f_{ori}$, the new FedRec $f_{unl}$ is expected to erase all contributions from $\mathcal{U}^{'}$.

Evaluating the effectiveness of unlearning is a big challenge.
Theoretically, retraining the federated recommender model from scratch after removing $\mathcal{U}^{'}$ can get an ideal model $f_{ret}$ that is guaranteed to completely forget the influence of $\mathcal{U}^{'}$.
But we cannot directly compare the model parameters between the unlearned model and the fully retrained model as there are too many stochastic processes during federated training.
Therefore, the common practice is to take $f_{ret}$ as the baseline model and evaluate the effectiveness and efficiency of the proposed unlearning approach in some attack scenarios~\cite{tarun2021fast,wu2022puma}. In this paper, we also follow this protocol. 
Specifically, we design an adversary to disturb the FedRec's training process via a group of malicious users.
Then, we assess and analyze \emph{(1) whether FRU can recover the destroyed FedRec} and \emph{(2) the recovery efficiency (i.e., time cost and space costs) compared with some baselines.}

\section{Federated Recommendation Unlearning (FRU)}
\subsection{Overview}
To fast unlearn target users, FRU calibrates all users' historical model updates and aggregates these updates to reconstruct FedRec models.
Algorithm~\ref{alg_fru} describes the overall work flow of FRU in a general Fedrec framework.
FRU contains two parts: efficient on-device update storing (Algorithm~\ref{alg_fru} Line 16, 19-20) and update revision for fast recovery (Algorithm~\ref{alg_fru} Line 3-5, 23-38).
The efficient on-device update storing aims to record each client's historical updates on its resource-constrained device.
It contains two components: a user-item mixed negative sampling method to reduce the size of updated parameters and an importance-based update filtering to dynamically clean less important historical updates.
Based on the stored historical updates, FRU can roll back the FedRec to the point when the deleted user/client joins the FedRec and then calibrate the historical updates from that time point to recover the FedRec.

In the following subsections, we present the details of FRU.
For clarity, we use $\mathcal{M}(V)_{k}^{t}$ and $\mathcal{M}(U)_{k}^{t}$ to represent the updates of global (or public) parameters (e.g. item embeddings) and local (or private) parameters (e.g. user embeddings) of user $u_{k}$ at $t$th global round.
$\mathcal{M}(U)_{k}^{t}$ is not submitted to the central server for privacy concerns, while $\mathcal{M}(V)_{k}^{t}$ is uploaded to the server and the server will update the global parameters by aggregating these clients' updates.

\subsection{Base Federated Recommenders}
To present the generalization ability of FRU, we choose the two most commonly used recommenders (NCF~\cite{he2017neuralcoll} and LightGCN~\cite{he2020lightgcn}) as our base models and train them with the most typical federated learning protocol~\cite{ammad2019federated}.

Neural collaborative filtering (NCF~\cite{he2017neuralcoll}) extends collaborative filtering (CF) by leveraging an $L$-layer feedforward network (FFN) to model the complex relationship between users and items.
\begin{equation}
  \label{eq_ncf}
  \hat{r_{ij}} = \sigma (h^\top FFN([\mathbf{u}_{i}, \mathbf{v}_{j}]))
\end{equation}
where $\mathbf{u}_{i}$ and $\mathbf{v}_{i}$ are user and item embeddings, respectively. $[\cdot]$ is concatenation operation.

As a graph-based recommender, LightGCN~\cite{he2020lightgcn} treats each user and item as a distinct node. 
The user-item interactions can be viewed as a bipartite graph.
User and item embeddings are learned and updated by propagating their neighbors' embeddings:
\begin{equation}
  \label{eq_lightgcn}
    \mathbf{u}_{i}^{l} = \sum\limits_{j\in \mathcal{N}_{u_{i}}}\frac{1}{\sqrt{\left| \mathcal{N}_{u_{i}} \right|} \sqrt{\left| \mathcal{N}_{v_{j}} \right|}}\mathbf{v}_{j}^{l-1}, 
    \mathbf{v}_{j}^{l} = \sum\limits_{i\in \mathcal{N}_{v_{j}}}\frac{1}{\sqrt{\left| \mathcal{N}_{v_{j}} \right|} \sqrt{\left| \mathcal{N}_{u_{i}} \right|}}\mathbf{u}_{i}^{l-1}
\end{equation}
where $l$ is the propagation layer in LightGCN. Note that under FedRec settings, the embeddings of neighbor users are not accessible. 
Therefore, each client can only use the local user-item interaction bipartite graph to update user and item embeddings.

The federated learning protocol is as follows.
The user embedding $\mathbf{u_{i}}$ is private, initialized and maintained at each client.
The item embeddings $\mathbf {V} $ and other global model parameters are initialized and sent from a central server.
After receiving item embeddings and other global parameters, each client combines its local user embedding $\mathbf{u_{i}}$ and these global parameters to form a local recommender model and then updates these model parameters on the local training data.
After local updating, each client sends the updated global parameters back to the server and keeps the updated user embedding $\mathbf{u_{i}}$ locally.
The server applies the average aggregation for these global parameters to gain the new global parameters.

\subsection{Efficient On-Device Update Storing}
Compared with retraining from scratch, FRU speeds up the reconstruction process by accurately rolling back the FedRec and then calibrating the stored model updates.
FRU takes advantage of storing historical model updates (i.e., the log) to avoid retraining from scratch. As mentioned in Section~\ref{sec:intro}, we choose to store historical model updates on local devices since centrally storing all clients' historical updates will incur unaffordable storage costs when the number of clients is large~\cite{wang2020next}. 
Therefore, how efficiently utilizing the limited storage space for the historical model updates  at each client device is the key.

Generally, the item embeddings dominate the size of a local recommender model in the FedRec. Based on this observation, we mainly focus on how to reduce the size of item embedding updates on each device. This paper proposes an importance-based update selection mechanism to store only important item embedding updates  and a user-item mixed semi-hard negative sampling method to reduce the number of negative instances at each client.

\subsubsection{\textbf{Importance-based Update Selection}}\label{sec:imp_up_pr}
Instead of storing the updates of the whole item embedding table on each client device,  the importance-based update selection mechanism only stores the embedding updates of the client's interacted items and the sampled negative items, which are a tiny portion of the whole item set. Then,  FRU further reduces the storage costs by ignoring non-significant updates. Intuitively, the influences of a client/user on different items' embeddings are different.
For example, some items are very popular, and their embeddings are updated by many users. 
In this case, a single user can only have a very limited impact on these items. 
Another example is that when an item's embedding is already well-trained, a user can only make an ignorable effect.
For these items whose embedding updates are so small at a client,  FRU will not store them to reduce the storage costs on the device.
Specifically, for a client/user, FRU only stores the top $\alpha$ proportion of item embedding updates based on their updates' significance $\lVert \mathcal{M} (\mathbf{v}_{i})_{k}^{t} \rVert$, where  $\mathcal{M}(\mathbf{v}_{i})_{k}^{t}=\mathbf{v}_{i}^{t}-\mathbf{v}_{i}^{t-1}$.  $\mathbf{v}_{i}^{t-1}$ is the received embedding of item $v_{i}$ from the central server, and $\mathbf{v}_{i}^{t}$ is its updated embedding on the local training data $\mathcal{D}_{k}$ of client $u_{k}$.

\subsubsection{\textbf{User-item Mixed Semi-hard Negative Sampling}}
As mentioned in Section~\ref{sec:imp_up_pr}, a client only needs to store the embedding updates of its associated items.
The associated items include the client's interacted items and  sampled negative items. In this part, we propose a more efficient negative sampling method to reduce the number of negative samples at each client without compromising the model's performance so that FRU can further alleviate the log storage space.

For each user, the traditional negative sampling method randomly selects a small portion of its non-interacted items as negative samples.
However, some sampled negative items may not be informative enough to optimize the recommendation model.
In a centralized recommendation system, some recent studies~\cite{yin2017sptf,wang2018neural} show that hard negatives contain more helpful information for model optimization and convergence. They proposed two methods to sample hard negatives based on user embeddings~\cite{yin2017sptf} and adversarial model~\cite{wang2018neural}.
However, these centralized negative sampling methods cannot be directly adopted in FedRecs.
Specifically, the adversarial method brings much extra computation overhead, and each user's local data is highly sparse and biased.
Sampling hard negatives based on user embedding is also unreliable in FedRecs because only a small portion of users are selected to train at each global epoch, and  the frequency of updating user embeddings is lower than centralized recommenders.
Consequently, how to retrieve high-quality negative samples from the item pool is still challenging for FedRecs.

In this paper, we propose a user-item mixed semi-hard negative sampling strategy.
Specifically, we choose semi-hard negative samples from both user and item sides.
The sampling from the user side is the same as in traditional centralized recommenders~\cite{yin2017sptf}. We calculate the relevance between the user's embedding and each item's embedding in the candidate pool to choose hard negative samples for each client.
As mentioned before, the frequency of updating user embedding is lower than centralized recommenders, so the user embedding needs to take a relatively long time to be informative. To compensate for the unreliability of user embeddings at the early stage, we integrate sampling negative items from the item side.  Specifically,  we use the embedding centroid of the user's interacted items as the pseudo user embedding at the early model training stage, as item embeddings are updated more frequently than user embeddings in FedRecs to be more reliable at a very early stage of training. We adopt the element-wise average to calculate the item embedding centroid.  Then, we choose hard negative items by calculating the relevance score with the centroid vector.

We first select the top $2R$\% items as the candidate item pool based on the user-item mixed sampling strategy.
Then, we randomly select $N*\beta$ ($0<\beta<1$) negative samples from the candidate item pool to form semi-hard negative samples, and 
$N$ is the original size of negative samples for each user in the traditional negative sampling methods. We do not select the top negative items from the candidate pool (i.e., hard negatives) to avoid sample false negatives. The sampling mechanism can be formally described as follows:
\begin{equation}
  \label{eq_ui_mixed_sample}
  \begin{aligned}
    &\mathcal{V}_{k}^{u} = \mathop{argmax}\limits_{v_{i} \notin  \mathcal{D}_{k} \land \left| \mathcal{V}_{k}^{u} \right| = R\% * \left| \mathcal{V} \right|} R(\mathbf{u}_{k}^{t-1}, \mathbf{v}_{i}^{t-1}) \\
    &\mathcal{V}_{k}^{v} = \mathop{argmax}\limits_{v_{i} \notin  \mathcal{D}_{k} \land \left| \mathcal{V}_{k}^{v} \right| = R\% * \left| V \right|} R(\mathbf{v}_{k,cet}^{t-1}, \mathbf{v}_{i}^{t-1})\\
    &\mathcal{V}_{k}^{neg} = \mathop{Random}\limits_{\left| \mathcal{V}_{k}^{neg} \right| =N * \beta}(\mathcal{V}_{k}^{u} \cup \mathcal{V}_{k}^{v})
  \end{aligned}
\end{equation}
where $\mathbf{v}_{k,cet}^{t-1}$ is the average embedding of $u_{k}$'s interacted items at the beginning of epoch $t$. $R(\cdot)$ is the relevance measure function. Here, we use Euclidean distance to measure the relevance.
$N*\beta$ is the number of negative samples we finally select. 
The ablation study in the experiment shows that our proposed negative sampling method achieves comparable performance even with $\beta=0.5$.
 Therefore, our efficient sampling mechanism significantly reduces the number of required negative samples for each client, thus leading to less storage space for item embedding updates. 

\subsubsection{\textbf{Storage Space Cost Analysis}}\label{sec:store_analyze}
$\left| \mathcal{V} \right|$ is the size of total items, $\left| \mathcal{V}_{k}^{pos} \right|$ and $\left| \mathcal{V}_{k}^{neg} \right|$ are the sizes of positive items and selected negative items for user $u_{k}$.
Assume the number of global epochs in the FedRec is $B$, and  $b$\% users will be selected for model training at each global epoch.
So averagely, a user will be trained $B\times b\%$ times in the whole training process.
The cost for storing an item's embedding updates is $C$.
As mentioned previously, for a local recommender, the cost of storing model updates mainly happens in item embeddings.
Therefore, we focus on analyzing the space cost of storing item embedding updates here.
Before applying our efficient on-device update storing method, each client keeps the whole item embedding table's updates after training.
Averagely, the cost of storing update logs for a user $u_{k}$ is $b\% \times B \times \left| \mathcal{V} \right| \times C$.
By applying our proposed Importance-based Update Selection method, the storage space cost can be reduced to $b\% \times B \times \alpha(\left| \mathcal{V}_{k}^{pos} \right| + \left| \mathcal{V}_{k}^{neg} \right|)  \times C$. Then, by applying our proposed efficient negative sampling method, the storage space cost for each client can be further reduced to $b\% \times B \times \alpha(\left| \mathcal{V}_{k}^{pos} \right| + \beta\left| \mathcal{V}_{k}^{neg} \right|)  \times C$. Generally, in most FedRecs, the negative items are sampled with a certain ratio of the positive items. Assume the ratio is $1:n$, then, the cost of our FRU's storage for each client is $b\% \times B \times \alpha(1+ \beta n)\left| \mathcal{V}_{k}^{pos} \right|  \times C$.

Take the setting of our experiment on Steam-200k as an example, where $B=200$, $b=10$, $\alpha=0.5$, $\beta=0.5$. 
The average interacted item size $\left| \mathcal{V}_{k}^{pos} \right|$ is about $30$.
The negative sample ratio $n$ for NCF and LightGCN is $4$ and $1$, respectively.
On average, the space cost of storing model updates on each client device is about $900C$ for NCF and $450C$ for LightGCN.
The total item size is $5134$. 
As a result, each client only needs to pay an extra $17.5\%$ space cost when using NCF ($8.75\%$ when using LightGCN) compared with the FedRec without unlearning ability.

\begin{algorithm}[!ht]
  \renewcommand{\algorithmicrequire}{\textbf{Input:}}
  \renewcommand{\algorithmicensure}{\textbf{Output:}}
  \caption{FRU (\textbf{F}ederated \textbf{R}ecommendation \textbf{U}nlearning)} \label{alg_fru}
  \begin{algorithmic}[1]
    \Require global epoch $T$; local epoch $L$; speed-up factor $\lambda$; embedding size $e$; learning rate $lr$, \dots;
    \Ensure global parameter $\mathbf{V}$, local client embedding $\mathbf{u}_{k}|_{k \in \mathcal{U}}$;
    \State Initializing global parameter $\mathbf{V}_{0}$;
    \For {each round t = 0, 1, ..., $T$}
      \If {request unlearning users $\mathcal{U}^{'}$}
        \State $\mathbf{V}_{t}\leftarrow$\Call{Unlearning}{$t, \mathcal{U}^{'}, \mathbf{V}_{0}$};
        \State $t\leftarrow t-1$;
      \Else
        \State sampling a fraction of clients $\mathcal{U}_{t}$;
        \For{$u_{k}\in \mathcal{U}_{t}$ }
          \State $\mathcal{M}(V)_{k}^{t+1}\leftarrow$ \Call{ClientUpdate}{$u_{k},\mathbf{V}_{t}, L$};
        \EndFor
        \State $\mathbf{V}_{t+1}\leftarrow \mathbf{V}_{t} + agg(\mathcal{M}(V)_{k}^{t+1})$;
      \EndIf
    \EndFor
    \Function{ClientUpdate} {$u_{k},\mathbf{V}_{t}, L$}
      \State downloading $\mathbf{V}_{t}$ from the server;
      \State $\mathcal{V}_{k}^{neg}\leftarrow$ sampling negative items using E.q.~\ref{eq_ui_mixed_sample};
      \State $\mathcal{M}(U)_{k}^{t+1}, \mathcal{M}(V)_{k}^{t+1} \leftarrow$ training $L$ epochs on $\mathcal{V}_{k}^{neg} \cup \mathcal{V}_{k}^{pos}$;
      \State $\mathbf{u}_{k}^{t+1} = \mathbf{u}_{k}^{t} + \mathcal{M}(U)_{k}^{t+1}$;
      \State $\mathcal{M}(V)_{k}^{t+1}\leftarrow$ selecting updates based on $\lVert \mathcal{M} (V)_{k}^{t+1} \rVert$;
      \State storing $\mathcal{M}(V)_{k}^{t+1}$ and user embedding $\mathbf{u}_{k}^{t+1}$ locally;
      \State \Return $\mathcal{M}(V)_{k}^{t+1}$
    \EndFunction
    \Function{Unlearning} {T$, \mathcal{U}^{'}, \mathbf{V}_{0}$}
      \State enquiring $\mathcal{M}(V)_{k}^{1}$ from clients;
      \State $\bar{\mathbf{V}}_{1}\leftarrow \mathbf{V}_{0} + agg(\mathcal{M}(V)_{k}^{1}|_{u_{k}\notin\mathcal{U}^{'}})$;
      \For{t=2, ..., T}
        \For{$u_{k}\in \mathcal{U}_{t}/ \mathcal{U}^{'}$}
          \State $\hat{\mathcal{M}}(V)^{t+1}_{k}\leftarrow$ \Call{ClientUnlearning} {$u_{k},\bar{\mathbf{V}}_{t-1}, \lambda, L$};
        \EndFor
        \State enquiring $\mathcal{M}(V)_{k}^{t}$ from clients;
        \State $\bar{\mathcal{M}}(V)^{t}\leftarrow$ calculate with E.q.~\ref{eq_calibrating_new};
        \State $\bar{\mathbf{V}}_{t}\leftarrow$ global parameter update with E.q.~\ref{eq_update_global};
      \EndFor
      \State \Return $\bar{\mathbf{V}}_{T}$
    \EndFunction
    \Function{ClientUnlearning} {$u_{k},\bar{\mathbf{V}}_{t}, \lambda, L$}
      \State $\hat{\mathcal{M}}(V)^{t+1}_{k}\leftarrow$ \Call{ClientUpdate} {$u_{k},\bar{\mathbf{V}}_{t}, \lambda * L$};
      \State \Return $\hat{\mathcal{M}}(V)^{t+1}_{k}$
    \EndFunction
    \end{algorithmic}
\end{algorithm}

\subsection{Unlearning with Updates Revision}

When a group of users leave the FedRec service and request to forget their information at a certain time $t$, FRU will first roll back the federated recommender model to the initial state (i.e. $t=0$) or the state when the first one of these users joined the federated training process, and then calibrate all historical model updates on the remaining clients.
The basic idea of calibration is that we only calibrate the historical updates' direction while keeping the length of updates unchanged since the direction guides the model to fit the training data, which has been polluted by the removed users~\cite{nasr2019comprehensive}.
Our unlearning method is based on FedEraser~\cite{liu2021federaser}, which is a unlearning method for classification tasks.
But we calibrate updates based on the efficient on-device update storing since the number of clients are greatly larger than classification tasks, meanwhile, we consider how to revise private parameters (i.e. user embeddings) which only exist in Fedrec tasks.
In what follows, we will present how to perform unlearning in FRU, which is corresponding to function \textsc{Unlearning} and \textsc{ClientUnlearning} in Algorithm~\ref{alg_fru}.
To avoid complex presentation, in this part, we directly use $\mathcal{M}(V)_{k}^{t}$ to represent the model updates at time $t$ that are stored with our proposed efficient on-device update storing method. We use $\mathbf{V}$ to directly represent the global parameters since the item embedding table dominates the global parameters' part.

\subsubsection{\textbf{Search Calibrated Direction}} 
At $t$'th global training round, a group of users are selected to train the FedRec. We denote this group of users as $\mathcal{U}_{t}$.
We first remove the target users $\mathcal{U}^{'}$ who request to forget their information.
Then, for each remained user $u_{k} \in \mathcal{U}_{t}/\mathcal{U}^{'}$, we run  $\lambda * L$ local training epochs based on the  unlearned global model $\bar{\mathbf{V}}_{t-1}$ achieved in the last round and the local user embedding $\mathbf{u}_{k}^{t}$ to get new global parameters' updates $\hat{\mathcal{M}}(V)^{t}_{k}$ and new user embedding updates $\hat{\mathcal{M}}(U)^{t}_{k}$.
Note that the users selected to recover the FedRec are the same as in the original training at round $t$, except the removed users.
Here, $L$ is the original local training epochs, and $\lambda$ is a \emph{speed-up factor}. 
We need much fewer local training epochs because we only want to approximate the update direction that can make the model fit the remaining data at this step.
It is worth mentioning that the global model $\mathbf{V}_{0}$ is the initial global recommender model on the central server that has yet received any update from clients, therefore, when $t = 1$, we can directly get the unlearned global model through aggregating updates from the remaining users: $\bar{\mathbf{V}}_{1}= \mathbf{V}_{0} + agg(\mathcal{M}(V)_{k}^{1}|_{u_{k}\notin\mathcal{U}^{'}})$.
The above operations can be seen in Algorithm~\ref{alg_fru} Line 25, 37.

\subsubsection{\textbf{Aggregate and Modify Updates}}
After the first step, the client computed new global parameter updates $\hat{\mathcal{M}}(V)^{t}_{k}$ and user embedding updates $\hat{\mathcal{M}}(U)^{t}_{k}$.
For the global parameter updates, clients at first upload their new computed updates $\hat{\mathcal{M}}(V)^{t}_{k}$ and the stored updates $\mathcal{M}(V)_{k}^{t}$ to the central server.
Then, the central server aggregates these updates and combines the new updates' direction with the original updates' length to construct the new calibrated updates $\bar{\mathcal{M}}(V)_{k}^{t}$, since the original updates' direction guides model to fit old training data.
\begin{equation}
  \label{eq_calibrating_new}
  \bar{\mathcal{M}}(V)^{t} = \left\| agg(\mathcal{M}(V)_{k}^{t})|_{u_{k}\notin\mathcal{U}^{'}} \right\| \frac{agg(\hat{\mathcal{M}}(V)^{t}_{k})|_{u_{k}\notin\mathcal{U}^{'}}}{\left\| agg(\hat{\mathcal{M}}(V)^{t}_{k})|_{u_{k}\notin\mathcal{U}^{'}} \right\|}
\end{equation}
$agg(\cdot)$ is aggregation strategy.

Then, the global parameters can be recovered at the central server based on the above calibrated updates, as follows:
\begin{equation}
  \label{eq_update_global}
  \bar{\mathbf{V}}_{t} = \bar{\mathbf{V}}_{t-1} + \bar{\mathcal{M}}(V)^{t}
\end{equation}

For user embedding updates, we directly use $\hat{\mathcal{M}}(U)^{t}_{k}$ to update user embedding: $\bar{\mathbf{u}}_{k}^{t} = \bar{\mathbf{u}}_{k}^{t-1} + \hat{\mathcal{M}}(U)^{t}_{k}$,  because the updating frequency of user embeddings is much lower than item embeddings, and the previous user embedding updates may not be reliable.

FRU repeats the above process at each global round to achieve the unlearned model.

\subsubsection{\textbf{Time Complexity Analysis}} 
One of the key advantages of FRU is that it can accelerate the reconstruction of the FedRec, compared with retraining from scratch.
The  speed-up ratio is mainly related to the speed-up factor $\lambda$, which allows clients to perform fewer rounds of local training.
Note that the local training time costs dominate the time complexity of the whole federated unlearning process. 
In our experiments, we set $\lambda=0.1$.  Under this setting, FRU can achieve up to 10x speedup compared with retraining from scratch.  
Empirically, FRU is $7$x faster than retraining from scratch.

\section{Experiments}
\subsection{Experimental Settings}~\label{sec_exp_setup}
\textbf{Datasets.}
We adopt two commonly used datasets for the federated recommendation: MovieLens-100k\footnote{https://grouplens.org/datasets/movielens/100k/}~\cite{harper2015movielens} and Steam-200k~\cite{cheuque2019recommender}.
MovieLens-100k contains 100,000 interactions, 943 users, and 1,682 movies, while Steam-200k includes 3,753 users, 5,134 steam games, and 114,713 user-game interactions.
Following~\cite{he2017neural,zhang2022pipattack,rong2022fedrecattack}, all the observed interactions are converted to $r_{ij}=1$. 
$80\%$ data and $20\%$ data are treated as train and test set.

\noindent\textbf{Baselines.} 
As there are no federated unlearning works for the on-device recommendation, we construct the following two baselines: \textbf{Retrain} and \textbf{FedRemove}.
Retrain means deleting the target users that we need to forget and then retraining the federated recommender model from scratch with the remaining users.  
We take 'Retrain' as the reference method that can lead to ideal unlearning results (e.g., ground-truth results). FedRemove is the method that simply removes the target users' global parameter updates and directly aggregates the remaining clients' updates during each global round when recovering the FedRec. In other words, compared with FRU, FedRemove does not include calibrating operations. Therefore, its recovery speed  is very fast.

\noindent\textbf{Evaluation Metrics.}
We employ the widely used Hit Ratio at rank $10$ (HR@10) and Normalized Discounted Cumulative Gain at rank 10 (NDCG@10) to measure the recommendation performance.

Evaluating the effectiveness of the federated  unlearning is not an easy task, as analyzed in Section~\ref{sec:intro}.
In this paper, we evaluate the unlearning effectiveness in the following attack scenario.
We assume a group of users are compromised. 
The malicious users attempt to perturb the FedRec's training process by uploading poisoning gradients to the server.
Specifically, the malicious user/client uploads flipped gradients to the server.
Simply flipping the gradients cannot severely destroy the FedRec, as many normal users may counteract the flipped gradients with correct gradients.
Therefore, we further add slight noise in the flipped gradients and use a randomly generated factor $\gamma$ to control the flipped gradients' magnitude.
\begin{equation}
  \label{eq_attack}
  \widetilde{\mathcal{M}}(V)_{k}^{t} = -\gamma\mathcal{M}(V)_{k}^{t} + \mu
\end{equation}
where $\widetilde{\mathcal{M}}(V)_{k}^{t}$ is poisoning updates generated by malicious user $u_{k}$.
$\gamma$ is a random factor sampled from a uniform distribution.
$\mu$ is a noise sampled from a normal distribution where the mean and standard deviation are calculated based on $\mathcal{M}(V)_{k}^{t}$.

To erase the influence of these malicious clients/users, we first delete these users from the FedRec and then apply our proposed FRU and the two baselines to recover the FedRec. In this scenario, we only need to evaluate whether the malicious clients' poisoning influence can be erased.

\noindent\textbf{Parameter Settings}
For both NCF, the dimension of user and item embedding is $64$. 
We adopt $4$  neural layers with dimensions $128$, $256$, $128$, and $64$  to process the concatenated user and item embeddings.
For LightGCN, the dimension of user and item embedding is $64$. 
We use $1$ layer for the graph propagation. 
The original negative sampling ratio is  $1:4$ and $1:1$ for NCF and LightGCN, following previous studies~\cite{zhang2022pipattack,ammad2019federated,he2020lightgcn}.
All other hyperparameters for these two models are the same.
Specifically, the learning rate, batch size, local epoch, and the maximum global epoch are set to $0.001$, $64$, $20$, and $200$, respectively.
$\alpha$, $\beta$, and $\lambda$ are $0.5$, $0.5$, $0.1$.

\begin{table*}[!ht]
  \centering
  \caption{Comparison of recovering attacked federated recommenders. ``IUS'' is short for importance-based update selection. ``Attacked'' shows the federated recommender's performance destroyed by the attack method. ``Retrain'' train Fedrec on the data without malicious users, whose performance can be referred as the performance of FedRec that has not been attacked.}\label{tb_main_result}
  \renewcommand{\arraystretch}{0.9}
  \begin{tabular}{ll|cccc|cccc}
  \hline
  \multirow{2}{*}{}                      & \multicolumn{1}{c|}{\multirow{2}{*}{}} & \multicolumn{4}{c|}{\textbf{MovieLens-100k}}                                                                                                 & \multicolumn{4}{c}{\textbf{Steam-200k}}                                                                                               \\ \cline{3-10} 
                                         & \multicolumn{1}{c|}{}                  & \multicolumn{2}{l}{\textbf{malicious 10\%}}                          & \multicolumn{2}{l|}{\textbf{malicious 20\%}}                          & \multicolumn{2}{l}{\textbf{malicious 10\%}}                       & \multicolumn{2}{l}{\textbf{malicious 20\%}}                       \\ \hline
  \textbf{Base Model}                    & \textbf{Method}                        & \multicolumn{1}{l}{\textbf{hit}} & \multicolumn{1}{l}{\textbf{ndcg}} & \multicolumn{1}{l}{\textbf{hit}} & \multicolumn{1}{l|}{\textbf{ndcg}} & \textbf{hit}                    & \textbf{ndcg}                   & \textbf{hit}                    & \textbf{ndcg}                   \\ \hline
  \multirow{5}{*}{\textbf{Fed-NCF}}      & \textbf{Attacked}                  & 55.34                            & 30.10                             & 49.02                            & 25.15                              & 81.05                           & 47.58                           & 76.28                           & 44.34                           \\
                                         & \textbf{FedRemove}                     & 51.06                            & 24.12                             & 45.28                            & 28.22                              & 76.77                           & 45.33                           & 49.30                           & 27.72                           \\
                                         & \textbf{Retrain}                       & {\ul 57.24}                      & {\ul 31.33}                       & {\ul 58.33}                      & {\ul 31.84}                        & {\ul 85.64}                     & {\ul 49.72}                     & {\ul 87.05}                     & {\ul 49.73}                     \\
                                         & \textbf{FRU w/o IUS}                   & 57.96                            & 30.57                             & 58.61                            & \textbf{32.77}                     & 88.39                           & 50.76                           & 91.19                           & 51.87                           \\
                                         & \textbf{FRU}                           & \textbf{58.43}                   & \textbf{31.26}                    & \textbf{60.28}                   & 32.38                              & \textbf{88.42}                  & \textbf{50.86}                  & \textbf{91.76}                  & \textbf{52.33}                           \\ \hline
  \multirow{5}{*}{\textbf{Fed-LightGCN}} & \textbf{Attacked}                  & 59.91                            & 29.92                             & 45.32                            & 24.19                              & \multicolumn{1}{c}{88.56}       & \multicolumn{1}{c}{51.95}       & \multicolumn{1}{c}{86.52}       & \multicolumn{1}{c}{50.84}       \\
                                         & \textbf{FedRemove}                     & 57.48                                & 27.17                                 & 56.67                            & 31.41                              & \multicolumn{1}{c}{87.22}       & \multicolumn{1}{c}{50.68}       & \multicolumn{1}{c}{90.14}       & \multicolumn{1}{c}{51.75}       \\
                                         & \textbf{Retrain}                       & {\ul 61.99}                      & {\ul 32.31}                       & {\ul 62.22}                      & {\ul 32.15}                        & \multicolumn{1}{c}{{\ul 92.14}} & \multicolumn{1}{c}{{\ul 53.85}} & \multicolumn{1}{c}{{\ul 93.64}} & \multicolumn{1}{c}{{\ul 54.31}} \\
                                         & \textbf{FRU w/o IUS}                   & 62.47                            & 33.25                             & 62.50                            & \textbf{33.29}                     & \multicolumn{1}{c}{92.07}       & \multicolumn{1}{c}{53.12}       & \multicolumn{1}{c}{93.26}       & \multicolumn{1}{c}{53.85}       \\
                                         & \textbf{FRU}                           & \textbf{63.90}                   & \textbf{34.61}                    & \textbf{63.06}                   & 31.53                              & \multicolumn{1}{c}{\textbf{93.57}}       & \multicolumn{1}{c}{\textbf{55.07}}       & \multicolumn{1}{c}{\textbf{93.79}}       & \multicolumn{1}{c}{\textbf{54.28}}       \\ \hline
  \end{tabular}
  \end{table*}
\subsection{Main Results and Analysis}
In this part, we present and discuss the main experimental results on erasing the poisoning influence from malicious users.
We evaluate the performance of our proposed FRU from two aspects: (1) whether the FedRec is recovered; (2) the efficiency of unlearning (i.e., recovering the FedRec).

\subsubsection{\textbf{Removing Malicious Users' Influence}}\label{sec:effect_unl}
Table~\ref{tb_main_result} shows the comparison of recommendation accuracy. 
``Attacked'' means that the FedRecs are attacked by malicious users.
``Retrain'' removes malicious users and then retrains FedRecs from scratch.
Table~\ref{tb_main_result} shows that FRU can recover the attacked FedRecs and achieve even better performance than Retrain in all cases, demonstrating FRU's effective and generic unlearning ability for different FedRec models.
FedRemove cannot work well with NCF or LightGCN, indicating the importance of calibrating updates.
In addition, with importance-based update selection, FRU can perform even better.
This observation indicates that only important updates should be calibrated, and over-calibration can negatively affect model performance.

\subsubsection{\textbf{Efficiency of Unlearning}}\label{sec:eff_unl}
\begin{figure}[!ht]
  \centering
  \includegraphics[width=0.46\textwidth]{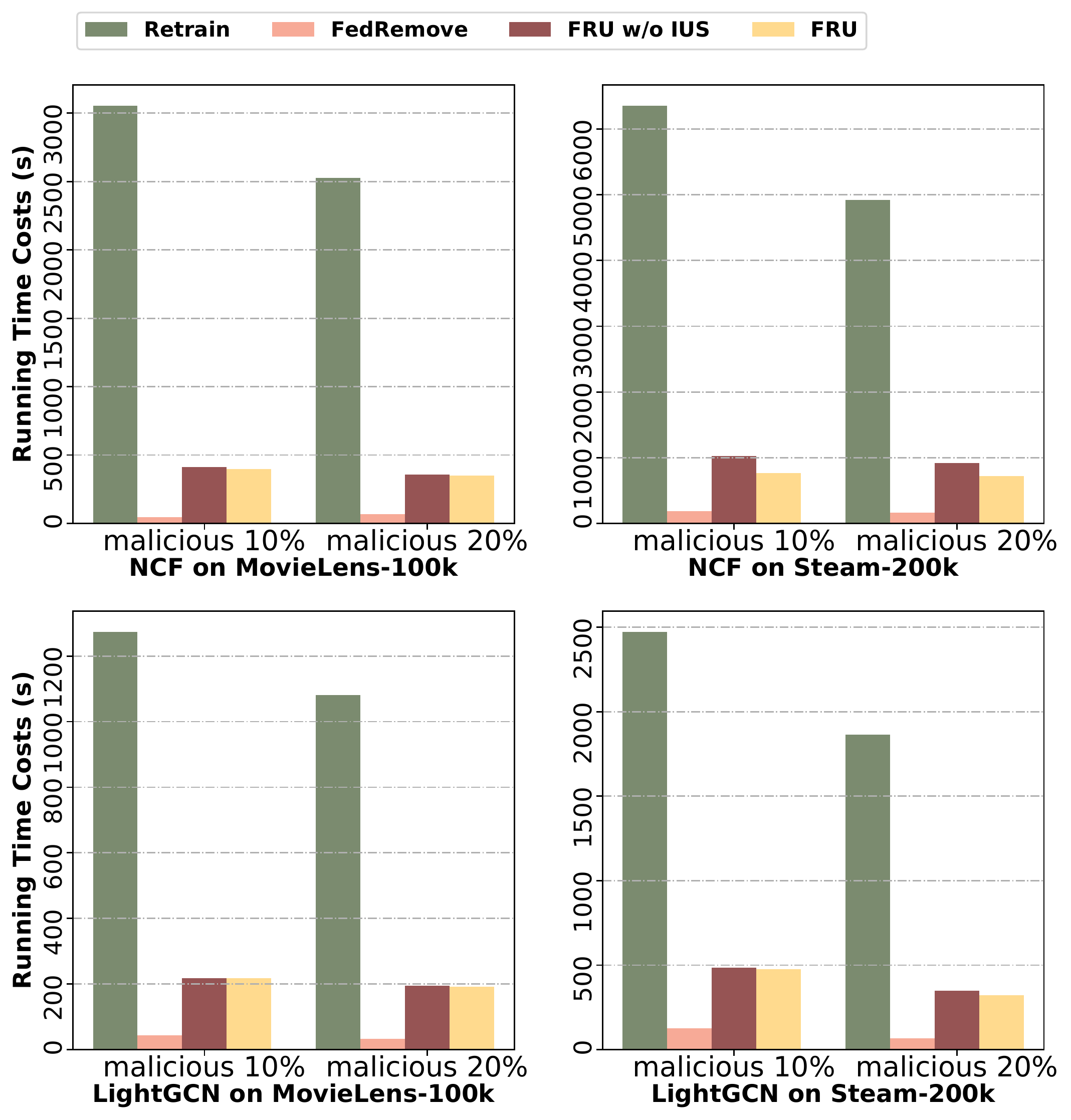}
  \caption{Time comparison of recovering FedRecs on MovieLens-100k and Steam-200k.}\label{fig_time_cost}
\end{figure}
Compared with retraining from scratch, one of the most important advantages of FRU is that it can speed up FedRec's recovery.
Fig.~\ref{fig_time_cost} shows the time costs of different unlearning methods on different datasets.
Under our settings of FRU (i.e. $\lambda=0.1$), FRU is $7$x faster than Retrain for both NCF and LightGCN models.
FRU is also a little bit faster than FRU w/o IUS because each client in FRU has fewer item embedding updates to calibrate. 
As expected, FedRemove is the most efficient unlearning method as it does not need to calibrate updates. However, its unlearning effectiveness is the worst, as shown in Table~\ref{tb_main_result}.

\subsection{Ablation Study and Hyperparameter Analysis}
In this part, we investigate $\alpha$, $\beta$, and $\lambda$'s influence, respectively.
Experiments are conducted on MovieLens-100k and Steam-200k with NCF and LightGCN, but we only present the results on MovieLens-100k due to the space limitation, and similar results and trends are observed on Steam-200k.
When investigating one component, the hyperparameters in other components keep unchanged.
\begin{figure}[!ht]
  \centering
  \includegraphics[width=0.47\textwidth]{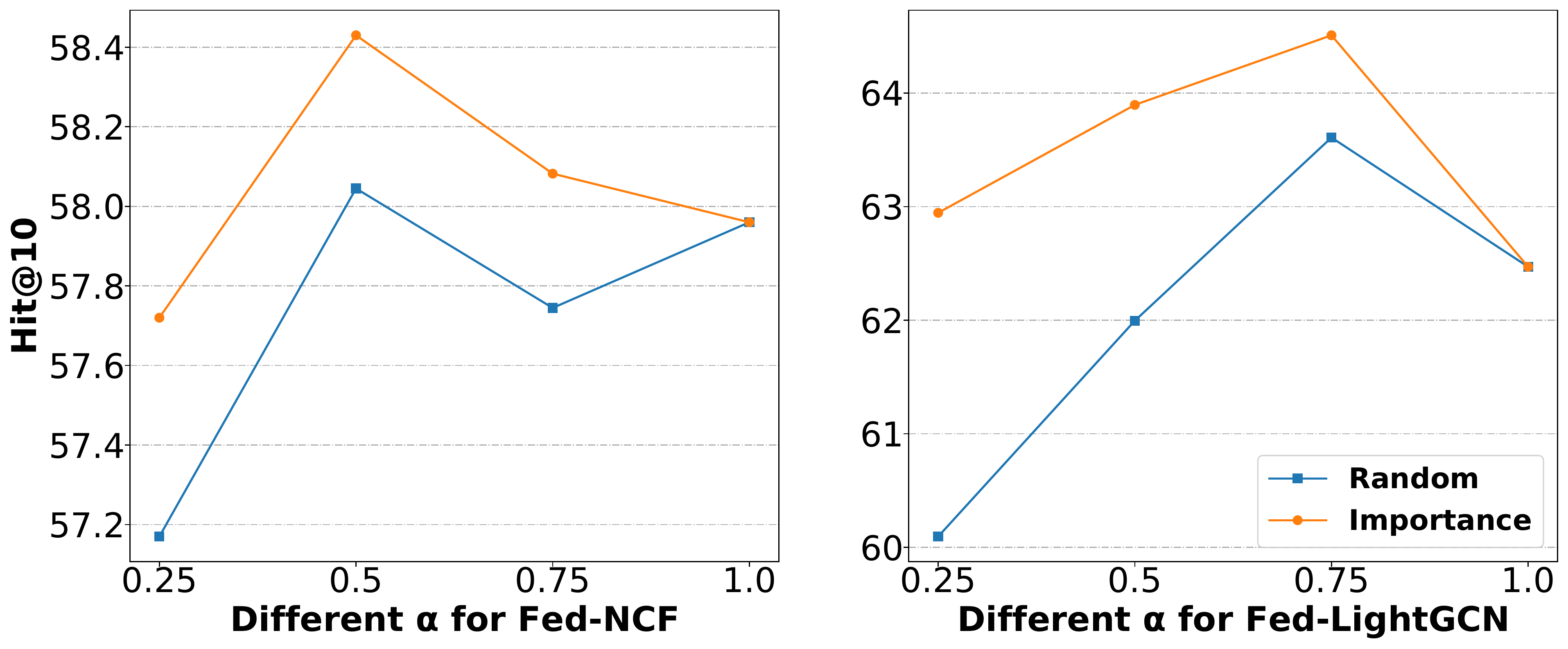}
  \caption{Different values of $\alpha$ for NCF and LightGCN on MovieLens-100k when recovering FedRecs from 10\% malicious users' attack. ``Random'' (blue line) means randomly selecting updates to store, while ``Importance'' (orange line) selects updates based on their importance.}\label{fig_imp_iup}
\end{figure}
\subsubsection{\textbf{Effect of Importance-based Update Selection}}
We explore different values of $\alpha$ for importance-based update selection and compare the results with randomly selecting updates.
The results are shown in Fig.~\ref{fig_imp_iup}.
When $\alpha=1.0$, FRU degrades to FRU w/o IUS since all updates will be stored.
In Fig.~\ref{fig_imp_iup}, our importance-based selection method outperforms the random selection strategy with all different $\alpha$ values.
In addition, the trends also show that FRU's performance first increases to a peak point and then decreases with continually increasing $\alpha$ value (i.e., keeping more item embedding updates). This phenomenon indicates that storing and calibrating too many unimportant updates will hurt FRU's performance.

\subsubsection{\textbf{Effect of User-item Mixed Semi-hard Negative Sampling}}
  \begin{figure}[!ht]
    \centering
    \includegraphics[width=0.46\textwidth]{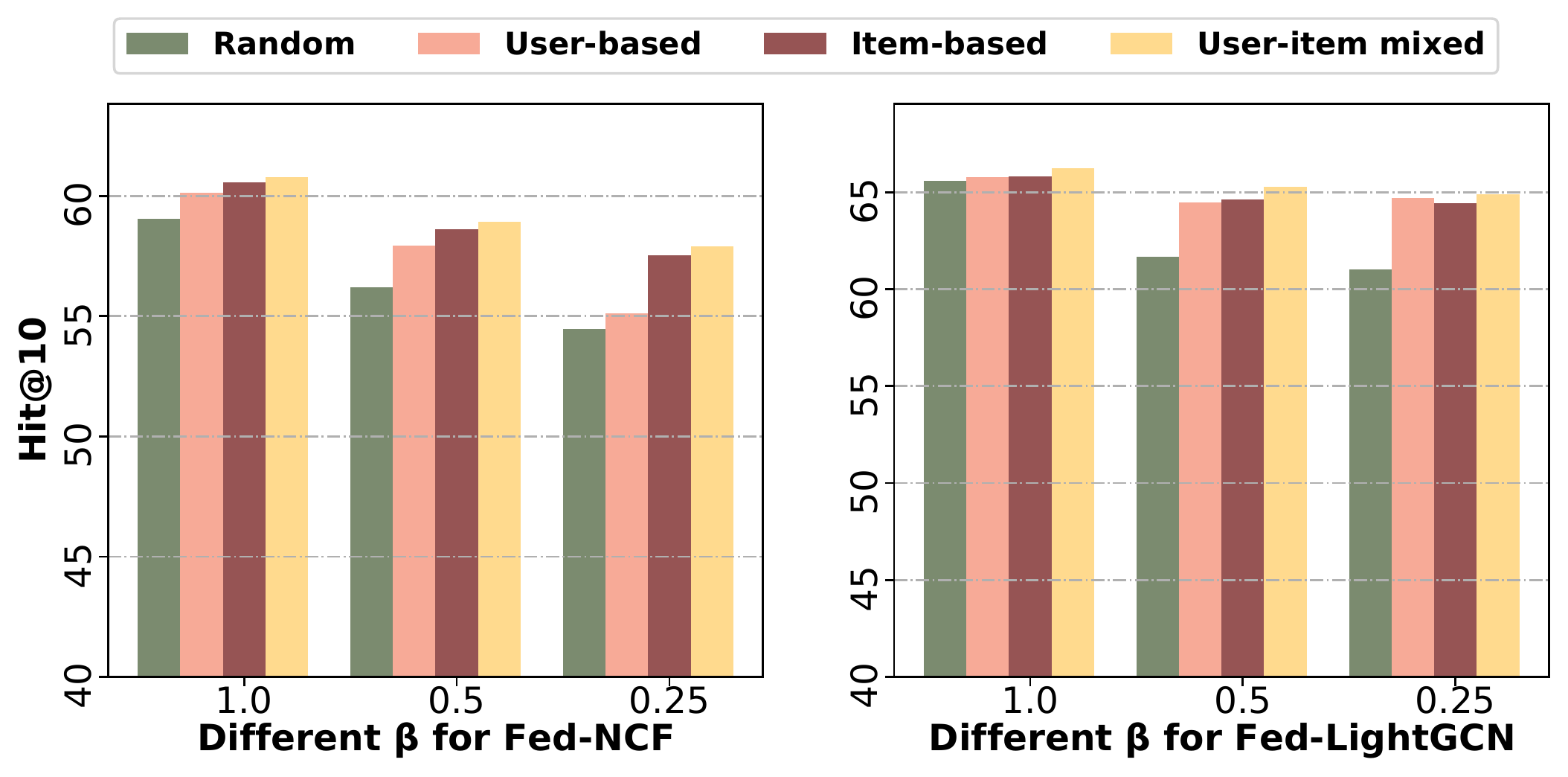}
    \caption{Different values of $\beta$ for NCF and LightGCN on MovieLens-100k without attacks. ``Random'' means randomly selecting negative samples, ``User-based'' and ``Item-based'' select negatives based on the relevance with the user embedding and the centroid of interacted items' embeddings, respectively.}\label{fig_neg_samp}
  \end{figure}
To analyze our proposed negative sampling method's effectiveness, we train FedRecs without attackers and explore three different values of $\beta$.
$\beta=1.0$ represents keeping the original sampling ratio in the traditional sampling method.
$\beta=0.5$ means our negative sampling rate is 50\% of the original one.
In Fig.~\ref{fig_neg_samp}, our user-item mixed semi-hard negative sampling method outperforms all baselines under the same $\beta$ values.
Another important observation is that our proposed sampling method with $\beta=0.5$ achieves almost the same performance as the random sampling method with $\beta=1.0$. As a result, using our proposed negative sampling method, we can reduce the negative sample size and further reduce the number of item embedding updates at each iteration without comprising the recommendation performance.

\subsubsection{\textbf{Impact of Speedup Factor}}
\begin{table}[!ht]
  \caption{The comparison of different $\lambda$ values to recover NCF and LightGCN destroyed by 10\% malicious users on MovieLens-100k.}\label{tb_lambda}
  \renewcommand{\arraystretch}{0.9}
  \begin{tabular}{l|cc|cc}
  \hline
                & \multicolumn{2}{c|}{\textbf{Fed-NCF}}                                           & \multicolumn{2}{c}{\textbf{Fed-LightGCN}}                                      \\ \cline{2-5} 
  $\mathbf{\lambda}$              & \multicolumn{1}{l}{\textbf{Hit}} & \multicolumn{1}{l|}{\textbf{Time Costs (s)}} & \multicolumn{1}{l}{\textbf{Hit}} & \multicolumn{1}{l}{\textbf{Time Costs (s)}} \\ \hline
  \textbf{0.1}  & \textbf{58.43}                   & \textbf{396.9}                               & 63.90                            & \textbf{217.3}                              \\
  \textbf{0.25} & 57.04                            & 894.6                                        & 61.28                            & 394.1                                       \\
  \textbf{0.5}  & 56.06                            & 1675.45                                      & \textbf{64.24}                   & 730.3                                       \\
  \textbf{0.75} & 53.44                            & 2270.61                                      & 62.27                            & 1029.2                                      \\ \hline
  \end{tabular}
  \end{table}
The speedup of recovering FedRecs mainly depends on the speedup factor $\lambda$.
In this part, we explore different values of $\lambda$ to show the unlearning recommendation results and the recovery time costs.
From Table~\ref{tb_lambda} we can observe that  when we increase $\lambda$, NCF's performance will drop.
For LightGCN, the relationship between the speedup factor and its performance is a little complex, but overall, it obtains lower performance with a larger $\lambda$ value.
For all FedRecs, the recovery time costs consistently increase with increasing $\lambda$ values.

  \section{Related Work} 
  \textbf{Federated Recommendation.} 
  Ammad et al.~\cite{ammad2019federated} propose the first general framework of FL with implicit feedback.
  Muhammad et al.~\cite{muhammad2020fedfast} ameliorate the user sampling and aggregation strategy to accelerate the convergency process.
  Besides utilizing implicit feedback, Lin et al~\cite{lin2020fedrec} and Liang et al.~\cite{liang2021fedrec++} attempt to exploit explicit feedback with FL.
  Kharitonov et al.~\cite{kharitonov2019federated} adopt online learning to learn online feedback.
  Wu et al.~\cite{wu2021fedgnn} incorporate GNN to improve federated recommendation performance further.
  Aside from improving the system's performance, many works also focus on attacking federated recommendations, such as ~\cite{zhang2022pipattack,rong2022fedrecattack,yi2022ua,wang2022fast}
  
  \noindent\textbf{Machine Unlearning.} 
  The work of machine unlearning can be classified into two categories: \emph{Exact Unlearning} and \emph{Approximate Unlearning}~\cite{nguyen2022survey}.
  For exact unlearning, methods are designed to theoretically guarantee that models can totally remove the influence from certain data.
  These methods usually require complicated mathematical calculations. Therefore, they can only be used for specific simple machine learning models~\cite{izzo2021approximate,baumhauer2020machine,brophy2021machine,schelter2021hedgecut,chen2019novel,ginart2019making}.
  More recently, to achieve unlearning in more complex models (e.g. deep learning), Bourtoule et al.~\cite{bourtoule2021machine} propose SISA. This method accelerates retraining through splitting the dataset into multiple partitions and training sub-models on these data shards.
  Based on SISA, some research takes steps into applying unlearning in deep learning models~\cite{chen2021graph}.
  However, all these methods have to utilize all data information, which is forbidden in FedRecs.
  
  Unlike exact unlearning, approximate unlearning methods relax the requirements of certifiably erasing data contributions. 
  Otherwise, it only provides a statistical assurance that the unlearned model forgets the targeted data to a large extent.
  Until now, 
  all federated learning-based unlearning approaches belong to approximate unlearning.
  For example, ~\cite{liu2020learn} and \cite{liu2021federaser} delete samples' influence via gradients.
  Wu et al.~\cite{wu2022federated} introduce knowledge distillation to unlearn federated models.
  Liu et al.~\cite{liu2022right} adopt the Newton methods to speed up retraining.
  However, most of these methods are applied to classification tasks. 
  Compared with FedRecs, these tasks have much fewer clients (users) and do not have private model parameters.
  Therefore, these methods cannot be directly used in FedRecs.
  
  \section{Conclusion}
  This paper proposes the first federated unlearning framework for on-device recommendation, called FRU. 
  Specifically, FRU stores all users' historical updates and reconstructs FedRecs by calibrating these updates, which is similar to the log-based rollback mechanism in database management systems.
  We propose an importance-based update selection strategy and a novel negative sampling method to efficiently store historical updates on resource-constrained devices.
  To evaluate FRU's unlearning ability, we propose an attack method that employs a group of compromised clients to perturb FedRecs' training process.
  Then, we apply FRU to erase these malicious users' influence.
  FRU is model-agnostic and can be incorporated in most FedRecs.
  We conduct experiments with two popular recommenders on two real-world recommendation datasets.
  The results show that FRU can eliminate specific users' influence and efficiently recover the FedRecs with $7$x  speedup.
  Finally, ablation studies are investigated to show the contributions of FRU's different components.

\begin{acks}
  This work is supported by Australian Research Council Future Fellowship (Grant No. FT210100624), Discovery Project (Grant No. DP190101985).  
  This work contributes to industry knowledge engineering in OpenTrek which is a technical brand of industry intelligence in Alibaba Cloud. 
\end{acks}

\bibliographystyle{ACM-Reference-Format}
\bibliography{sample-base}










\end{document}